\documentclass[submission,copyright,creativecommons]{eptcs}
\providecommand{\event}{FMAS 2022} 

\usepackage{iftex}
\usepackage{amsfonts,amsmath,amssymb,amsthm}
\usepackage{tikz}
\usepackage{multicol}
\usepackage{url}
\usepackage{graphicx}
\usepackage[utf8]{inputenc}
\usetikzlibrary{shapes,arrows,positioning,fit,calc,petri}

\ifpdf
  \usepackage{underscore}         
  \usepackage[T1]{fontenc}        
\else
  \usepackage{breakurl}           
\fi
\usepackage[english]{babel}

\definecolor{gray}{HTML}{666666}		
\definecolor{lightblue}{HTML}{006699}		
\definecolor{darkblue}{HTML}{003355}		
\definecolor{lightgreen}{HTML}{669900}		
\definecolor{bluegreen}{HTML}{33997e}		
\definecolor{orange}{HTML}{e2661a}		
\definecolor{purple}{HTML}{7d4793}		
\definecolor{commentscolor}{rgb}{0,0.6,0}
\definecolor{numbercolor}{rgb}{0.5,0.5,0.5}
\definecolor{stringcolor}{rgb}{0.2,0.6,0.8}
\definecolor{backcolor}{rgb}{0.93,0.93,0.93}
\definecolor{keywordcolor}{rgb}{0,0,0.75}
\definecolor{keywordcolorbis}{rgb}{0,0.6,0.5}
\definecolor{identifiercolor}{rgb}{0.25,0.25,0.25}

\usepackage{listings}
\lstdefinelanguage{rotboLanguage}{
    morekeywords = [1]{skillset},
    morekeywords = [2]{data, resource, event, skill},
    morekeywords = [3]{initial, guard, input, output, precondition, start, invariant, progress, update, interrupt, success, failure},
    morekeywords = [4]{period},
    otherkeywords = {!,>,<,.,;,-,=,not,and,true,false,0,1,2,3,4,5,6,7,8,9},
    morekeywords = [5]{!,-,>,<,=,.,0,1,2,3,4,5,6,7,8,9},
    morekeywords = [6]{not,and},
    morekeywords = [7]{true,false},
    morecomment = [l]{//},
    morecomment = [s]{/*}{*/},
    morecomment = [s]{/**}{*/},
    keywordstyle = [1]\color{lightblue}\bfseries,
    keywordstyle = [2]\color{lightblue}\bfseries,
    keywordstyle = [3]\color{lightblue}\bfseries,
    keywordstyle = [4]\color{lightblue}\ttfamily,
    keywordstyle = [5]\color{orange}\ttfamily,
    keywordstyle = [6]\color{lightblue}\ttfamily,
    keywordstyle = [7]\color{darkblue}\ttfamily,
    commentstyle = \color{lightgreen}
}
\lstdefinestyle{rotboLanguage}{
    language=rotboLanguage,
    basicstyle=\ttfamily\small\color{black},
    identifierstyle=\color{identifiercolor},
    commentstyle=\color{commentscolor},
    numberstyle=\scriptsize\color{numbercolor},
    stringstyle=\color{stringcolor},
    keywordstyle=\bfseries\color{keywordcolor},
    keywordstyle={[2]\bfseries\color{keywordcolorbis}},
    breakatwhitespace=true,
    breaklines=true,
    keepspaces=true,
    numbers=left,
    numbersep=0pt,
    showspaces=false,
    showstringspaces=false,
    showtabs=false,
    tabsize=2,
}

\lstdefinelanguage{smt2}{
    morekeywords = [1]{declare, datatypes, define, fun, const, assert},
    morekeywords = [2]{Bool, Int, Real},
    morekeywords = [3]{check,sat,get,model,eval},
    otherkeywords = {>,<,.,=,-,not,and,true,false,0,1,2,3,4,5,6,7,8,9},
    morekeywords = [4]{not,and,>,<,=,.,0,1,2,3,4,5,6,7,8,9},
    morekeywords = [5]{-},
    morekeywords = [6]{true,false},
    morecomment = [l]{;},
    keywordstyle = [1]\color{lightblue}\bfseries,
    keywordstyle = [2]\color{lightblue}\bfseries,
    keywordstyle = [3]\color{purple}\bfseries,
    keywordstyle = [4]\color{orange}\ttfamily,
    keywordstyle = [5]\color{lightblue}\ttfamily,
    keywordstyle = [6]\color{lightblue}\ttfamily,
    commentstyle = \color{lightgreen}
}

\lstdefinestyle{mySmt}{
    language=smt2,
    basicstyle=\ttfamily\small\color{black},
    identifierstyle=\color{identifiercolor},
    commentstyle=\color{commentscolor},
    numberstyle=\scriptsize\color{numbercolor},
    stringstyle=\color{stringcolor},
    keywordstyle=\bfseries\color{keywordcolor},
    keywordstyle={[2]\bfseries\color{keywordcolorbis}},
    breakatwhitespace=true,
    breaklines=true,
    keepspaces=true,
    numbersep=0pt,
    showspaces=false,
    showstringspaces=false,
    showtabs=false,
    tabsize=2,
}

\title{SkiNet\\ A Petri Net Generation Tool for the Verification of Skillset-based Autonomous Systems}
\author{Baptiste Pelletier
\institute{ONERA/DTIS\\ Université de Toulouse, France}
\institute{LIRMM \\
Univ. Montpellier, CNRS France}
\email{baptiste.pelletier@onera.fr}
\and
Charles Lesire \qquad\qquad David Doose
\institute{ONERA, DTIS\\ Toulouse, France}
\email{firstname.lastname@onera.fr}
\and
Karen Godary-Dejean \qquad\qquad Charles Dramé-Maigné
\institute{LIRMM, Univ. de Montpellier, CNRS\\ Montpellier, France}
\email{karen.godary-dejean@umontpellier.fr \qquad\qquad charles.drame-maigne@ens-paris-saclay.fr}
}
\def\titlerunning{SkiNet}
\def\authorrunning{B. Pelletier, C. Lesire, D. Doose, K. Godary-Dejean \& C. Dramé-Maigné}
\def\copyrightholders{B. Pelletier, C. Lesire, D. Doose, \\ K. Godary-Dejean \& C. Dramé-Maigné}
\begin{document}
\maketitle

\tikzstyle{decision} = [diamond, draw, fill=blue!20, 
    text width=4.5em, text badly centered, node distance=3cm, inner sep=0pt]
\tikzstyle{block} = [rectangle, draw, fill=blue!20, 
    text width=5em, text centered, rounded corners, minimum height=4em]
\tikzstyle{line} = [draw, -latex']
\tikzstyle{cloud} = [draw, ellipse,fill=red!20, node distance=3cm,
    minimum height=2em]

\begin{abstract}
The need for high-level autonomy and robustness of autonomous systems for missions in dynamic and remote environment has pushed developers to come up with new software architectures. A common architecture style is to summarize the capabilities of the robotic system into elementary actions, called skills, on top of which a skill management layer is implemented to structure, test and control the functional layer. However, current available verification tools only provide either mission-specific verification or verification on a model that does not replicate the actual execution of the system, which makes it difficult to ensure its robustness to unexpected events. To that end, a tool, SkiNet, has been developed to transform the skill-based architecture of a system into a Petri net modeling the state-machine behaviors of the skills and the resources they handle. The Petri net allows the use of model-checking, such as Linear Temporal Logic (LTL) or Computational Tree Logic (CTL), for the user to analyze and verify the model of the system.
\end{abstract}

\section{Introduction}

The use of autonomous systems has spread widely in the recent years, with applications in industrial automation, scientific exploration, rescue or environment monitoring missions. They are able of executing critical tasks and missions that need a high level of dependability.

For the sake of improving the trust in the autonomy of robotic systems, different validation and verification methods (V\&V) can be used, such as formal methods, like model-checking, to verify that user-defined properties will be guaranteed once the system is deployed. The verification of the system behavior whatever happens requires to predict all possible system configurations, failure points and their appropriate responses, as well as the complexity of a dynamic, sometimes unknown environment. For a system to answer each of these challenges, hardware and software complexity must increase, which makes the modelling and analysis processes more difficult. Using formal methods and verification tools such as model-checking, it is possible to prove the robustness of the system throughout its design stages.

Various software control architectures have been developed with always the same objectives in mind: making the system easier to specify, and its robustness easier to verify. Looking at the system with a higher-level of abstraction can make the verification process easier for users with few experience in middleware and low-level architecture. But abstraction becomes quickly limited when facing dynamic environments and complex tasks. An ideal model would then be one that is abstract enough to allow an easy specification, while also providing great verification possibilities, which is the goal of the present study.

The software architectures tackled in this work are skill-based architectures, also called task-based \cite{Lozano1983,Albore2021,Pedersen2016}, where the robotic system is decomposed into elementary actions. This approach usually comes with a high-level of abstraction of the software and hardware of the system, while offering a wide range of specification possibilities. A controller layer is then implemented, to test the architecture and control the system. The architecture used for this work will be the Skillset formalization of Albore et al. \cite{Albore2021}, which is composed of resources, skills and events and can model the robot as well as its environment and operator decisions. The strength of this formalization is that a code generating tool is provided, which will guarantee that the execution of the system will be as specified in the skillset.

The present work was conducted with the objective of improving the current verification capabilities of the skillset architecture by using a Petri net equivalent to the skillset model. The long-term objective is to offer an accessible model-checking tool for this architecture, to ensure that all people involved in the design of a robotic system can understand and contribute to the verification process with their own technical knowledge. Indeed, current autonomous systems are becoming more costly and multidisciplinary, while being used in remote and hazardous environments, where human assistance is nearly impossible. This requires scientific inputs during the design phase from researchers requesting valuable data, field experts that will put physical limitations to the system, and technical operators that will design the robotic hardware and software. In order for all these participants to contribute, the verification process must be simplified. For instance, formalisms such as Linear Temporal Logic (LTL) or Computation Tree Logic (CTL) have important capabilities to verify the behavior of a system, but demand a solid knowledge in the functioning of model-checking in order to use them properly.

Various formal methods are used for the purpose of property verification, and the present tool we present in this paper, SkiNet, uses Petri nets to translate the skillset model and perform model-checking indirectly. The end user never manipulates the Petri net nor the model-checking tools, as SkiNet performs all these steps in the background and only provides to the user the valuable informations for the skillset design process. This is especially useful as Petri nets modelling complex systems can become difficult to read visually, and properties to check also become proportionally more complex.


Thus, this paper presents SkiNet and the methods it uses. Section 2 will begin by looking at the related work tackling the verification of high-level software architectures using formal methods. Then, section 3 will sum up the background, with the definitions of Petri nets and the skillset architecture. Section 4 will go over the content of the generated Petri net and the generation process. Section 5 will prove the correctness of the generated Petri net with regards to the state-machine behavior of skills and resources. Finally, section 6 will show a few examples of model-checking done on the generated Petri net to verify the behavior of the skillset model, before moving on to the conclusion in section 7. An illustrating example with the Boston Dynamics Spot\textsuperscript{\tiny\textregistered} robot, shown in Fig. \ref{fig:spot_picture}, with the \texttt{spot} skillset, in Fig. \ref{fig:spot_skillset}, used for generating the controller code.
The SkiNet tool and instructions are available publicly for the readers to try it out themselves: \url{https://gitlab.com/onera-robot-skills/skinet-release}.

\section{Related works}

To facilitate the development of autonomous systems, skill-based architectures have been designed to decompose the system into elementary actions. An early take on the subject was done in \cite{Lozano1983}, with many new concepts emerging with an increasing demand for autonomous systems from industries \cite{Pedersen2016}. The main advantage of such architectures are their modularity, easy reconfiguration and repurposability, as well as making robotic programming easier for end users. The skills are then composed to create more complex tasks to achieve specific goals or missions, while using feedback from sensors and actuators \cite{Pedersen2016,Albore2021,Steinmetz2016,Schou2018,Lesire2018}. The mission design using skill-based architecture is often coupled with deliberative functions to create autonomous systems capable of adapting to their environment or faulty behaviors, often using model-based architectures \cite{Ingrand2017,Patra2021}. This is a growing need in domains such as underground, underwater \cite{Mcgann2008,Perdomo2010,Zereik2018} or space exploration \cite{Ingham2001,Ingham2006}. However, such functions imply a precise specification of all possible faults that can arise, for instance by designing fault-trees \cite{Hereau2021}, and implementing a proper risk-management architecture \cite{McGhan2015,Ayton2020}.

In order to prepare the autonomous system for high-risk missions and guarantee a-priori its robustness to the dynamic environment it will evolve in and to the faults that could arise, formal methods based on model-checking have been developed, sometimes at a very early development stage. Formal methods are used to tackle the verification of either the mission design or the entire robotic system. 
Ingrand \cite{Ingrand2019} gathers the state of the art of V\&V formal methods for autonomous systems. Albore \cite{Albore2021} shows how missions can be designed through Behavior Trees, allowing for robust mission specification and fault management, however such method restricts the verification of the system to a specific mission. 
Evans \cite{Evans2017} proposes a Model-Based Mission Assurance approach to improve the safety and robustness of a system in the early development stage, using Assurance models synthesis, implementing fault trees and Bayesian nets as inputs for SysML diagrams, opening interesting perspectives for the implementation of fault management into software architecture.
Nardone \cite{Nardone2019} proposes a methodology for the V\&V of satellite operational mode management specification, using mu-calculus logic. The method offers an interesting insight on the use of mu-calculus logic for V\&V. Gross \cite{Gross2017} also uses formal specification for the early stages of spacecraft design and attitude control system and shows the cost-reducing capabilities of using model-checking in a large-scale project. Finally, Louis \cite{Louis2017} designed a mission controller based on a Fault Management System for an underwater automated vehicle. This approach, while being very safe with regards to system integrity, also demands the inputs from experts to specify the fault model and calibrate the controller correctly.

Petri nets \cite{Peterson1977,Murata1989} are used in a wide range of applications such as robotics, industrial management or video-games \cite{Reuter2015}, with a wide range of tools readily available to manipulate and verify Petri nets, such as the Tina toolbox (Time Petri net Analyzer) \cite{Berthomieu2004,TinaWebsite}, and thus indirectly verify the modeled system or mission \cite{Lesire2018,Costelha2012,Mura2001}. Costelha \cite{Costelha2007,Costelha2012} uses Petri nets to model and verify robotic tasks. This early approach is very close to our goal here: a design-analysis-design approach that improves user experience when designing the robot model before using the real robot. However, the nets were manually built, while the present work is based on written specification synthesis.
Reza \cite{Reza2009} uses both Petri nets and converted Fault trees into nets for verification and safety analysis for systems based on Requirements State Machine Language (RSML). This approach is similar to the present work, with automatic synthesis of RSML into fault trees and Petri nets. A similar approach was proposed by Yan \cite{Yan2017} for autonomous mission reliability modelling. Kwon \cite{Kwon2004} uses specialized colored Petri nets to model context-aware agent-based applications. The systems are decomposed into context-independent patterns, which is an approach also considered by Figat \cite{Figat2022}, with hierarchical Petri nets, where layered patterns are used to describe components of multi-agent robotic systems. These approach show how much complexity can be put into Petri nets to model a system, and how easily customizable they are to suit the modelling need of every user and/or system. Multiple agents missions verification is also tackled by \cite{Palomeras2010}, and such approaches open future perspectives on the present work for the use of skill-architecture based Petri nets with multiple coordinating agents. Finally, Mahulea \cite{Mahulea2018} proposes an automatic generation of Petri nets for boolean based robot planning, with generation and verification processes also similar to the present work, but without the skill-based architecture context.

Finally, controller synthesis using formal methods has been widely developed to close the gap between a safe model and a controller code that respects this model, while relieving the programmers from writing their own middleware. Foughali proposes the use of GenoM3 \cite{Foughali2019}, a framework to specify robotic systems in the form of component-based timed transition systems, from which controller synthesis can be performed using templates. GenoM3 uses a lower-level approach than the skillset architecture presented in this work, but raises critical points on the usability of controller synthesis frameworks with other existing V\&V tools. Figat \cite{Figat2022} suggests hierarchical Petri nets to decompose multi-agents systems into functional layers. This hierarchical perspective allows for the use of ready-to-use templates to model low-level components, so that the user can focus on modelling and verifying the higher-level architecture. Controller synthesis is also performed using this framework, however the Petri nets are hand-written and not synthesized from user-written specification, which can make it difficult for a non-experienced user to manipulate. Lesire \cite{Lesire2018} creates a Skills Colored Petri net controller and synthesizes it into a ROS node to manage the execution of the system. Unfortunately, the resources system was not yet developed, which limited mission complexity. On top of this, Colored Petri nets are a special formalization of Petri nets, which limits the available tools for verification.

The present work uses the same framework as \cite{Albore2021}, where skillset specifications are written by the user, and a skill-manager layer is synthesized into a ROS node to complement with pre-built system control functions. Our work here is to provide a verification tool for this architecture, using Petri nets. The goal of the tool is to translate the user-written specification of the skillset into a Petri net, by using the definition of skill-nets presented in \cite{Lesire2018}, with the addition of the resources system and events. We also removed the colored net formalism to be able to use more verification tools. Controller code can be generated from the skillset specification, and our tool can be used for V\&V purpose, during both design and deployment phase, thus fully covering the aforementioned problematics of system development.

\section{Background}
\subsection{Petri net}

This section will sum up the definitions and notions of the generated Petri net, as defined by \cite{Peterson1977,Murata1989}, with the addition of transition priority as presented by Balbo \cite{Balbo2001}.

\theoremstyle{definition}
\newtheorem{definition}{Definition}[section]
\begin{definition}[Petri net]\label{def:DefPetriNet}
A Petri net $\left< N, m_0\right>$ is a tuple $N = (P,T,F)$ and an initial marking $m_0$, where:
\begin{itemize}
    \item $P$ and $T$ are two non-empty, disjoint and finite sets of places and transitions, respectively.
    \item $F \subseteq (P \times T) \cup (T \times P)$ a set of directed arcs. 
    \item $m_0 \in M$ the initial distribution of tokens, called the initial marking of the net, and $M = \{m_0,m_1,...,m_n\}$ the set of all possible markings of $N$.
\end{itemize}
This definition is extended with $N = (P,T,F,\succ)$, where $\succ$ is the priority relation, represented by directed arcs between transitions, with the source transition having a higher priority. This means that if two transitions $t_1, t_2 \in T$ are enabled, i.e. their input places have at least one token, and $t_1 \succ t_2$, then only $t_1$ is firable.
\end{definition}

For any transition $t \in T$, the sets of its input and output nodes are $^{\bullet}{t}$ and $t^{\bullet}$ respectively. While arcs are usually weighted in conventional Petri nets, only unitary arcs will appear in this paper, with an explanation in section 4.

Let $p \in P$ and $t \in T$ be a place and a transition. The marking of a place $p$ is noted $m[p]$. The firing of an enabled transition $t \in T$, with $\forall p \in ^{\bullet}{t},\ m[p] \geq 1$, leads to a new marking, or reachable state, $m'$. All the input places of $t$ loose a token, i.e. $\forall p \in ^{\bullet}{t},\ m'[p] = m[p] - 1$, and all the output places gain one token, i.e. $\forall p \in t^{\bullet},\ m'[p] = m[p] + 1$.

\subsection{Skillset architecture}

\begin{figure}[!ht]
    \centering
    \includegraphics[width=0.35\textwidth]{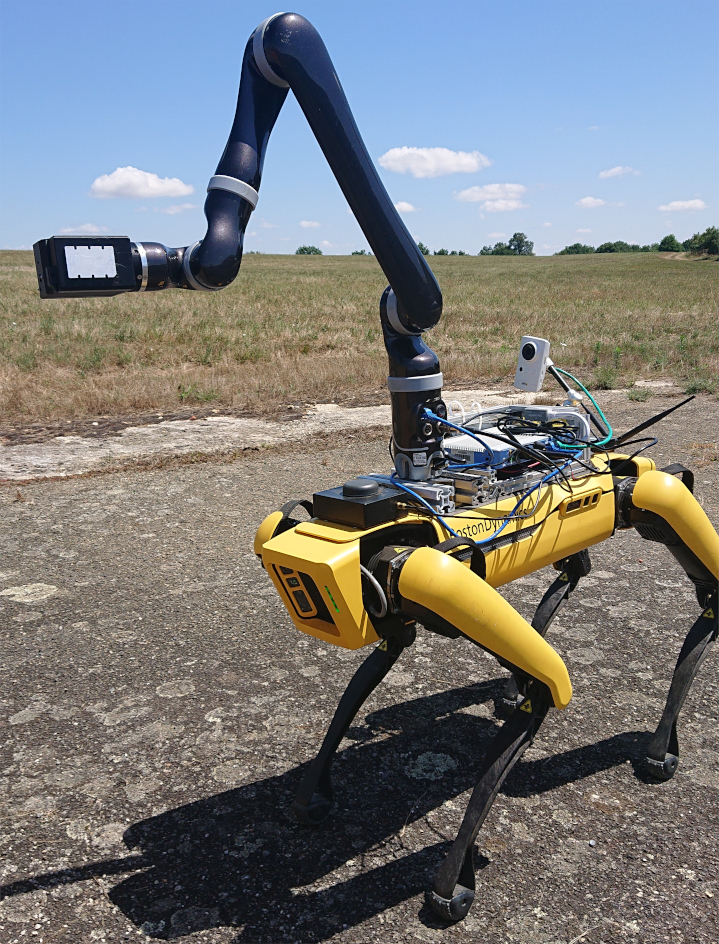}
    \caption{The Spot\textsuperscript{\tiny\textregistered} quadruped robot, mounted with extra payload. The controller code was generated from a skillset which was verified a-priori with SkiNet. The skillset in Fig \ref{fig:spot_skillset} is an extract of the actual running skillset, available at \url{https://gitlab.com/onera-robot-skills/skinet-release}.}
    \label{fig:spot_picture}
\end{figure}

\begin{figure}[!ht]
    \centering
    \includegraphics[width=0.8\textwidth]{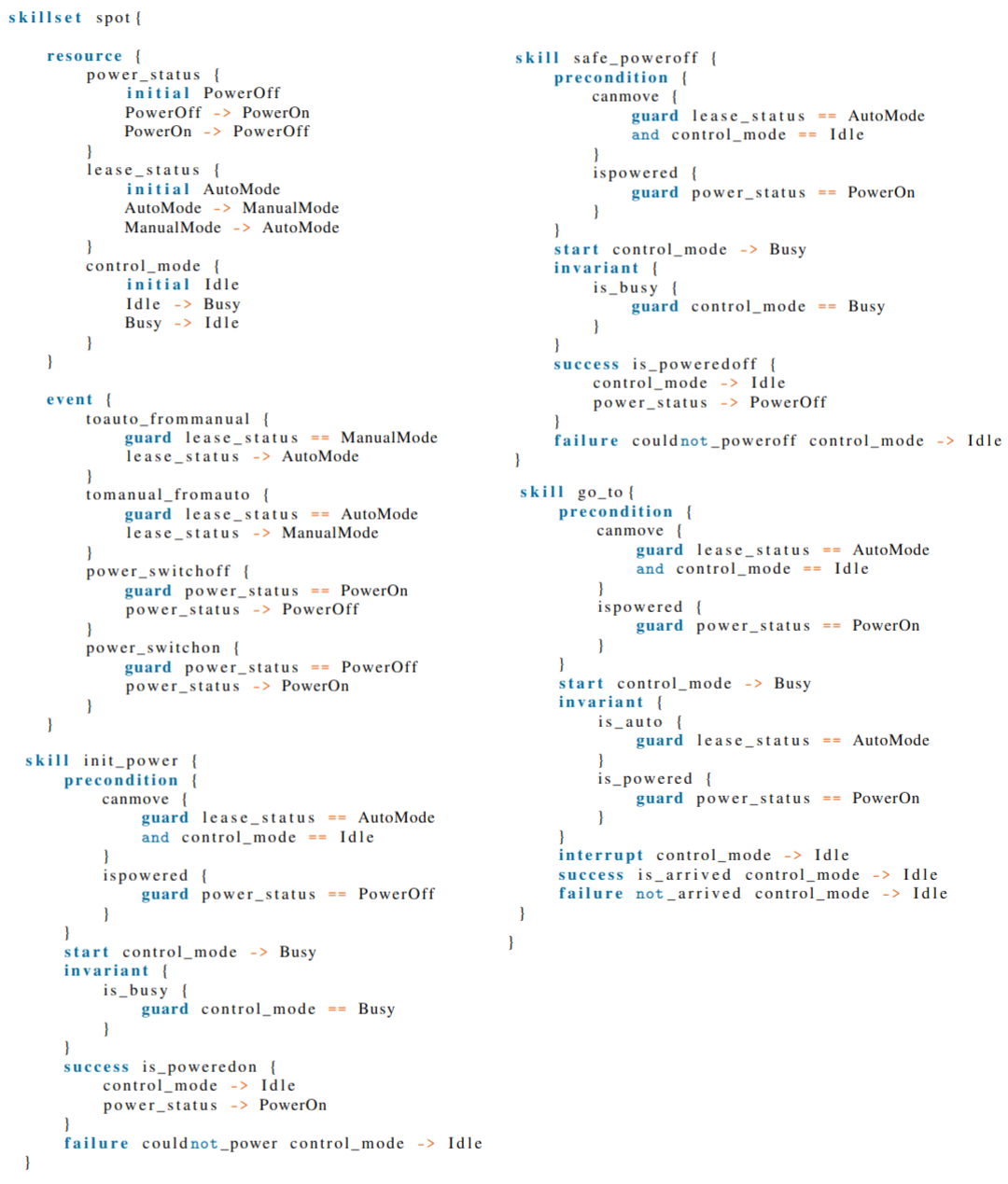}
    \caption{Extract of the \texttt{spot} skillset model for the Boston Dynamics Spot\textsuperscript{\tiny\textregistered} robot. Resources model motor power state, lease mode and control mode mutex. Events are used to represent the actions of the operator on the robot. Skills the robot can perform are \texttt{init_power} and \texttt{safe_poweroff} to start and cut motor power respectively, and a movement skill \texttt{go_to}.}
    \label{fig:spot_skillset}
\end{figure}

This section summarizes the elements of a Skillset used for the modelling and programming of autonomous systems, as defined by Albore \cite{Albore2021}, with some elements omitted as they are not used in this paper. The Skillset can represent both hardware and software elements of the system, and their interaction/execution. A tool called "robot language" generates C++ code based on the skillset specifications that follows this execution, with part of the execution code to be filled by the user, such as skills functions, exit conditions, events triggering, etc. More information on the skillset execution semantic can be found in \cite{Albore2021}. A skillset contains resources, resource guards and resource effects. These elements are assembled into skillset transitions, to create events and skills. We define all the elements of a skillset in the following section. An example skillset is given in \ref{fig:spot_skillset}, which was used for generating controller code for the Boston Dynamics Spot\textsuperscript{\tiny\textregistered} robot, a quadruped robot capable of carrying heavy payload and perform observation tasks, shown in Fig. \ref{fig:spot_picture}. This example showcases the syntax of the skillset specification language as it would be written by the user, which is significantly simpler than the underlying formal definition, given in the following section.

\subsubsection{Resources}

\begin{definition}[Resources]\label{def:DefResources}
    Resources are state-machines that represent the status of an element of the system (sensor, motorization status, mutex...). A resource $r$ is a tuple $(S^r,T^r)$, with a set of states $S^r = \{ S^r_0,...,S^r_{n-1}\}$, and a set of transitions between each state $T^r \subset (S^r \times S^r)$. A resource can only be in one of its states at any given time during the execution of a skillset, noted $state(r) \in S^r$, with $S^r_0$ the initial state.

    The set of states that can lead to a state $S^r_i$ and the set of states that can originate from $S^r_i$ are noted $^{\bullet}S^r_i$ and ${S^r_i}^\bullet$ respectively. The transition from $S^r_i$ to itself is a valid transition.
\end{definition}

\begin{definition}[Resources effects and guards] \label{def:DefResourcesEffectsGuards}
    Resource effects aim to change the state of a specific resource. An effect $\epsilon$ is a tuple $(r,S^r_i)$ of a resource and its next state. The origin state of the resource is not specified, as effects only mention the destination state. Effects can be empty and are only valid if they contain at most one effect for each resource.
    
    Resource guards are used to put conditions on the triggering of effects. A resource guard, or simply guard, is a logical formula $\phi : \{ S^r, r \in \mathcal{R} \} \to \{True,False\}$ on the states of the resources of $\mathcal{R}$.
\end{definition}

\subsubsection{Skillset transitions}

The skillset is then assembled by coupling guards and effects to create skillset transitions. Skillset transitions are used in events and skills, and represent the basis of the execution of the skillset. We also define solutions, the set of resource states that verify a resource guard.

\begin{definition}[Skillset transitions]
    A skillset transition is a tuple $\tau = ( \phi,\mathcal{E},\sigma ) $, formed with a guard $\phi$, a set of effect $\mathcal{E}$, and a state change $\sigma$. A state change is a tuple $\sigma = (state 1, state 2)$ which changes the state of a skill, later defined in Def \ref{def:DefSkillset}. We note $\mathcal{R}(\tau)$ the resources involved in either $\phi$ or $\mathcal{E}$.
    A skillset transition can be triggered if its guard is true. If triggered, its effects and state change are applied.

\end{definition}

\begin{definition}[Solutions]\label{def:solutions}
    $X_\tau$ is the set of all solutions of a skillset transition $\tau$. A solution is a set of resource states $x = \{x_r \in S^r, r \in \mathcal{R}(\tau) \}$ associated to a guard $\phi$, such that $\phi(x) = True$. $x$ only contains up to one state per resource.
    The solutions $X_\tau$ can be obtained with a boolean satisfaction problem solver (SAT solver).
\end{definition}

\subsubsection{Skillset}

\begin{definition}[Skillset]\label{def:DefSkillset}
A skillset is a tuple $\Sigma = (\mathcal{R},\mathcal{V},\mathcal{S})$ with:

\begin{itemize}
    \item $\mathcal{R}$ is a set of resources $\{r_1,r_2,...\}$.
    
    \item $\mathcal{V}$ is a set of events. An event is a skillset transition $\tau_\nu$ with state change $\sigma = \varnothing$. Events represent actions from the exterior of the system, such as from the operator or the environment, on the resources of the skillset. They can have a guard or none.
    
    \item $\mathcal{S}$ is a set of skills, which are elementary actions that can be executed by the system. A skill $s \in \mathcal{S}$ is a tuple $s = (name,pre,inv,state,\mathcal{T}_s)$, with:
    \begin{itemize}
        \item An identifier $name$.
        
        \item Preconditions $pre = \{pre_1, pre_2,...\}$ are a set of guards that must be verified to start the skill. If they are not satisfied while attempting to start the skill, we may have failure effects $\bar{pre}_i$ triggered on each unsatisfied guard.
        
        \item $inv = \{inv_1,inv_2,...\}$ is a set of guards, called invariants, that must remain true during the skill execution. An invariant failure results in an immediate termination of the skill and in the application of the failure effects $\bar{inv}_i$.
        
        \item A set of states $state$: idle $e_s$, running $i_s$ or terminated $x_{s,k}$, with the termination mode $k \in \{ \bar{pre},\bar{inv},succ,fail,int\}$. This notation is adapted from the colored Petri net formalism used by Lesire et al. \cite{Lesire2018}. At the initial state of the skillset, all skills are idle.
        
        
        \item A set of skillset transitions $\mathcal{T}_s = \{\tau_{s,start}, \tau_{s,\bar{pre},i},  \tau_{s,\bar{inv},i}, \tau_{s,succ}, \tau_{s,fail}, \tau_{s,int}\}$, with:
        \begin{itemize}
            \item Start $\tau_{s,start} = \{pre,start,e_s \to i_s\}$, with guards $pre = pre_1 \land pre_2 \land ...$, start effects $start$ and state change $e_s \to i_s$. $\tau_{s,start}$ is triggered when a skill is started and all preconditions $pre$ are verified. The $start$ effects are then applied and the skill goes from idle $e_s$ to running $i_s$.
            \item Precondition failures $\tau_{s,\bar{pre},i} = \{
            \neg pre_i,\bar{pre}_i,e_s \to x_{s,\bar{pre},i}\}$ are triggered if the start of a skill is attempted but the precondition $pre_i$ is not verified, i.e. resources states do not verify the resource guard.
            \item Invariant failures $\tau_{s,\bar{inv},i} = \{
            \neg inv_i,\bar{inv}_i,i_s \to x_{s,\bar{inv},i}\}$ are triggered if the skill is running and the invariant $inv_i$ is not verified.
            \item Success $\tau_{s,succ} = \{
            inv,succ,i_s \to x_{s,succ}\}$, failure $\tau_{s,fail} = \{
            inv,fail,i_s \to x_{s,fail}\}$ and interrupt $\tau_{s,int} = \{
            inv,int,i_s \to x_{s,int}\}$, are triggered when the invariants $inv$ are still verified and the skill execution ends with a success, failure or interrupt respectively.
        \end{itemize}
    \end{itemize}
    
    More detail on the skill execution state-machine can be found in \cite{Albore2021}. Note that multiple successes and failures can be added, but only one interrupt is possible.
    
\end{itemize}
\end{definition}

Using the \texttt{spot} skillset example from Fig. \ref{fig:spot_skillset}, the execution of a skill is as follows: in order to perform the \texttt{go_to} skill, the preconditions \texttt{canmove} and \texttt{ispowered} must be satisfied. In that case, starting the skill will trigger the $start$ effects. Upon starting the skill, invariants satisfaction is checked. While the guards \texttt{is_auto} and \texttt{is_powered} are respected, the skill is running, until the skill ends, triggering the associated effects $int$, $succ$ or $fail$.

\section{SkiNet Architecture}

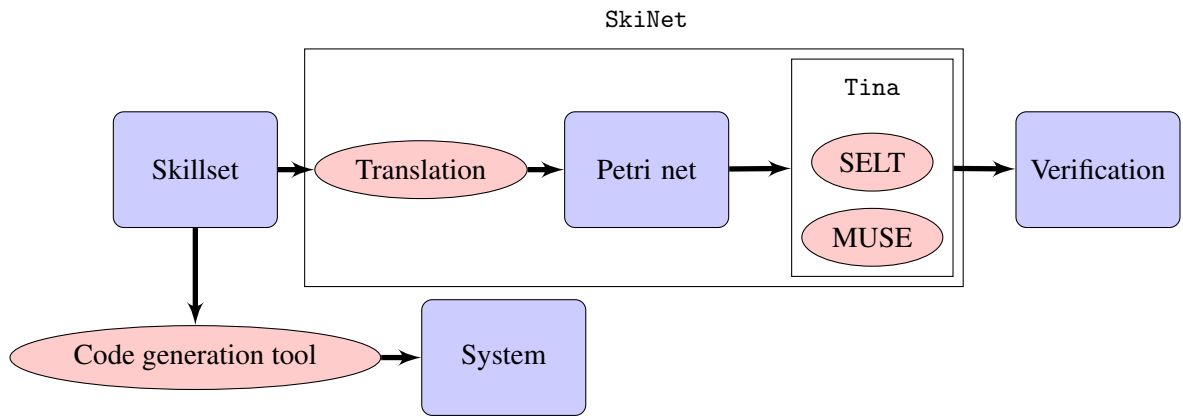
\begin{figure}[!ht]
    \centering
    \begin{tikzpicture}[node distance = 3.0cm, auto,title/.style={font=\fontsize{10}{10}\color{black!100}\ttfamily}]

        \node [block] (skillset) {Skillset};
        \node [cloud, right of=skillset] (translation) {Translation};
        \node [block, right of=translation] (net) {Petri net};
        \node [title, right of=net,yshift=1.1cm] (tina) {Tina};
        \node [cloud, below of=tina,node distance = 1cm] (selt) {SELT};
        \node [cloud, below of=selt,node distance = 1cm] (muse) {MUSE};
        \node [draw=black!100, fit={(selt) (muse) (tina)}] (tinabox){};
        \node [block, right of=tina,yshift=-1.1cm] (verif) {Verification};
        \node [title, above of=net,yshift=-1cm] (skinet) {SkiNet};
        \node [draw=black!100, fit={(translation) (net) (tinabox)}] (skinetbox){};
        \node [cloud, below of=skillset, yshift=0.5cm] (robotlang) {Code generation tool};
        \node [block, right of=robotlang,xshift=1.1cm] (robot) {System};
        
        \path [line,line width=2pt] (skillset) -- (translation);
        \path [line,line width=2pt] (translation) -- (net);
        \path [line,line width=2pt] (net) -- (tinabox);
        \path [line,line width=2pt] (tinabox) -- (verif);
        \path [line,line width=2pt] (skillset) -- (robotlang);
        \path [line,line width=2pt] (robotlang) -- (robot);
        
    \end{tikzpicture}
    \caption{SkiNet is a tool used to generate a Petri net from a skillset, on which model-checking can be performed, using tools taken from Tina \cite{Berthomieu2004,TinaWebsite}. SkiNet acts as a layer to avoid manipulating the tools directly. The verified skillset can then be used for controller code generation.}
    \label{fig:skinet}
\end{figure}

SkiNet was designed with the aim to translate the skillset of a robotic system into an equivalent Petri net that would reproduce its behavior, in order to perform model-checking using Petri net verification tools, and verify the behavior of the skillset indirectly. A graphical representation of SkiNet in its context is available in Fig. \ref{fig:skinet}. In this section, the translation process of a skillset $\Sigma=(\mathcal{R},\mathcal{V},\mathcal{S})$ into a Petri net $N = (P,T,F,\succ)$ will be explained, while section 5 will prove the correctness of the translation, and section 6 will show the verification capabilities of the tool using the generated Petri net.

In this work, we use the foundations of skill Petri nets presented by Lesire et al. \cite{Lesire2018}. However, the colored Petri net formalism used has been removed. While it allowed the net to be more expressive, colored Petri nets don't have the same range of verification tools as classical Petri nets, such as the Tina toolbox \cite{TinaWebsite}. The expressivity lost doing so is not relevant here, thanks to SkiNet which extracts the useful informations from the generated Petri net without the need for the user to look at it. Moreover, colored Petri nets only make the nets more compact, but do not reduce the size of the marking space $M$, so the complexity is identical. 

\subsection{Places of the generated Petri net}

The places of the generated Petri net are $P = P_r \cup P_s$, with:
\begin{itemize}
    \item $P_r = \{p^r_i, 0 \leq i \leq |S^r|, r\in \mathcal{R}\}$ the resource state places, with a place $p^r_i$ for each state $S^r_i \in S^r$ of each resource $r$ of the skillset $\Sigma$.
    \item $P_s = \{p^s_e,p^s_i,p^s_{x,k},k \in \{ \bar{pre},\bar{inv},succ,fail,int\}, s\in \mathcal{S} \}$ for $\mathcal{S} \in \Sigma$, the places representing the skills execution state. This representation is inspired from the work of Lesire et al. \cite{Lesire2018} and defined as follows:
    \begin{itemize}
        \item $p^s_e$ is the entry place. This place must be marked, i.e. $m[p^s_e] \geq 1$ for the skill $s$ to start.
        \item $p^s_i$ is the intermediate place. The start of the execution of the skill $s$ is represented by the start transition which will move the token from $p^s_e$ to $p^s_i$.
        \item $p^s_{x,k}$ are the exit places. Upon skill termination, the token is moved from $p^s_i$ to the place $p^s_{x,k}$ of the corresponding termination mode $k$.
    \end{itemize}
\end{itemize}

\subsection{Transitions of the generated Petri net}

For each skillset transition $\tau = (\phi,\mathcal{E},\sigma)$, a set of transitions $T_\tau$ is generated in the Petri net $N$.
For each solution $x \in X_\tau$ of $\phi$, a transition $t_x \in T_\tau$ is generated, if and only if:
\begin{itemize}

    \item The input places of $t_x$ are the resource state places in $x$: for each $x_r \in x$, we have $p^r_i$ the Petri net place of the resource state $S_i^r$, and $p^r_i \in\ ^{\bullet}t_x$. Up to one resource state place $p^r_i$ exists in $^{\bullet}t_x$ for each resource, as only one state per resource exists in each solution $x$, cf Def. \ref{def:solutions}.
    
    \item The output places of $t_x$ are the destination states $S^r_i$ of the effects $\epsilon \in \mathcal{E}$: for $S^r_i \in \epsilon, \epsilon \in \mathcal{E}$, we note $p^r_i$ the Petri net place of $S^r_i$, and $p^r_i \in\ t_x^{\bullet}$. Because there is at most one effect $\epsilon$ for each resource,cf Def. \ref{def:DefResourcesEffectsGuards}, then there is only at most one resource state place $p^r_i$ for each resource.
    
    \item The input and outputs are extended with the states in $\sigma = (state1, state2)$. We note $p^s_1$ and $p^s_2$ the skill state places of each state in $\sigma$, and we have: $\forall t_x \in T_\tau, p^s_1 \in\ ^{\bullet}t_x$ and $p^s_2 \in t_x^{\bullet}$. If $\sigma = \varnothing$, then no place is added to the inputs and outputs of $t_x$.
    
\end{itemize}

The generation of the transitions needs to be further complexified in order to overcome the limits of Petri nets. Indeed, there are cases of skillset transitions $\tau$ where a resource is evaluated by the guard $\phi$ but not present in the effects $\mathcal{E}$, and vice-versa. Therefore, we need to generate the transitions according to these issues:
\begin{itemize}
    \item If a resource is \emph{nominal}, i.e. both guarded and affected in $\tau$, then nothing needs to be done on $T_\tau$. The generated transitions simply move the tokens between the states of the resource as already specified in $\tau$.
    \item If a resource is \emph{unaffected}, i.e. guarded but with no effects in $\tau$, then the corresponding state place $p^r_x \in ^{\bullet}t_x$ would lose a token upon firing $t_x$, and this token would not be restituted. This would mean that the resource could be in an empty state, which is impossible. Therefore the token must be returned to the same state after firing $t_x$, i.e. $t_x^{\bullet} \leftarrow t_x^{\bullet} + p^r_x $.
    \item If a resource is \emph{unguarded}, i.e. affected but not guarded in $\tau$, then the corresponding destination state place $p^r_y \in t_x$ would receive a token, without first taking one from the state places of the resource. This would mean that the resource could be in two states at the same time, which is also impossible. Therefore, the token must be taken somewhere among the state places $p^r_i$ of the resource. However, we cannot anticipate where the token is for a given marking, therefore all possible resource state changes must be computed, given the transition exists among the valid transitions of $T^r$ that lead to $S^r_y$, i.e. $^{\bullet}S^r_y$, cf Def. \ref{def:DefResources}. We note $X_r$ the set of the state places $p^r_i$ corresponding to the resource states in $^{\bullet}S^r_y$, and we extend $X_\tau$ as:
    $X_\tau \leftarrow X_\tau \cup X_r$.
\end{itemize}

After building all the transitions, we have $\bigcup_{\tau} T_\tau$ the set of all transitions as defined previously. To this set, we add a reset transitions set $T_{reset}$, to allow for the repeatability of skills, which moves the token from the exit places $p^s_{x,k}$ to the entry place $p^s_e$, with one reset transition per exit place, i.e.:
\begin{equation}
    \forall s \in \mathcal{S}, \exists t_{s,reset,k} \in T_{reset}, ^{\bullet}t_{s,reset,k} = p^s_{x,k}
\end{equation}
\begin{equation}
    t_{s,reset,k}^{\bullet} = p^s_e, k \in \{ \bar{pre},\bar{inv},succ,fail,int\}
\end{equation}

The final set of transitions in the generated Petri net is then $T = ( \bigcup_{\tau} T_\tau )\cup T_{reset}$.

Finally, transitions of invariant failures $T_{\bar{inv}} = \bigcup T_{\tau,\bar{inv}}$ have a higher priority than all other transitions, i.e.:
\begin{equation}
   \forall t_{\bar{inv}} \in T_{\bar{inv}}, \forall t \in T - T_{\bar{inv}}: t_{\bar{inv}} \succ t
\end{equation}
This allows to respect the execution semantic of the skills, and forces a skill to terminate first if its invariant is broken before firing another transition, which could potentially "repair" the invariant.

\subsection{Examples with the \texttt{spot} skillset model}\label{sec:spot_example}

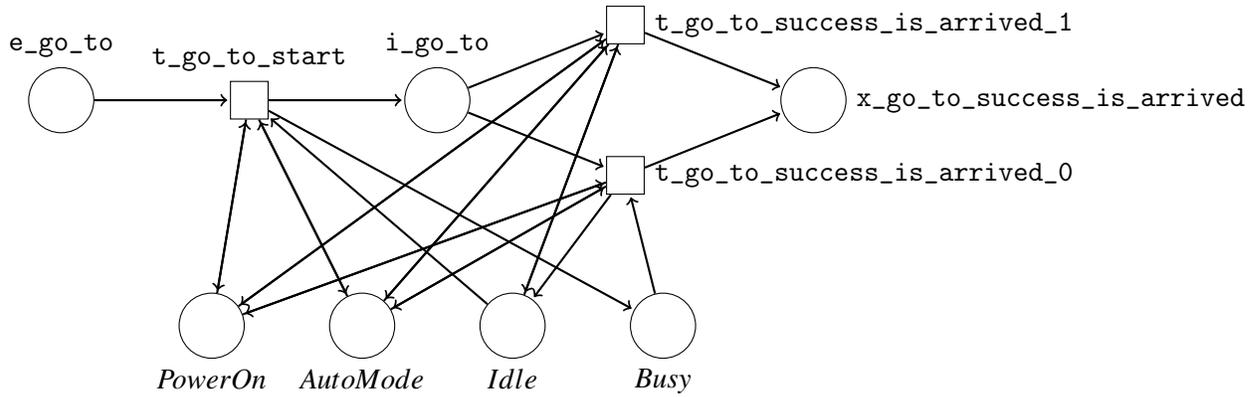
\begin{figure}[!ht]
    \centering
    \begin{tikzpicture}[node distance = 1cm, auto,title/.style={font=\fontsize{10}{10}\color{black!100}\ttfamily}]

    \node[place,label=below:$PowerOn$,tokens=0] (poweron) at (2,-3) {};
    
    \node[place,label=below:$AutoMode$,tokens=0] (auto) at (4,-3) {};

    \node[place,label=below:$Idle$,tokens=0] (idle) at (6,-3) {};
    \node[place,label=below:$Busy$,tokens=0] (busy) at (8,-3) {};

    \node[place,label=above:\texttt{e_go_to},tokens=0] (es) at (0,0) {};
    \node[place,label=above:\texttt{i_go_to},tokens=0] (is) at (5,0) {};
    \node[place,label=right:\texttt{x_go_to_success_is_arrived},tokens=0] (xs) at (10,0) {};

    \node[transition,minimum size=0.5cm,label=above:\texttt{t_go_to_start}] (tstart) at (2.5,0) {};
    \draw[thick] (poweron) edge[post] (tstart)
        (auto) edge[post] (tstart)
        (idle) edge[post] (tstart)
        (es) edge[post] (tstart)
        (tstart) edge[post] (is)
        (tstart) edge[post] (poweron)
        (tstart) edge[post] (auto)
        (tstart) edge[post] (busy);

    \node[transition,minimum size=0.5cm,label=right:\texttt{t_go_to_success_is_arrived_0}] (tsuccess0) at (7.5,-1) {};
    \draw[thick] 
        (busy) edge[post] (tsuccess0)
        (auto) edge[post] (tsuccess0)
        (poweron) edge[post] (tsuccess0)
        (is) edge[post] (tsuccess0)
        (tsuccess0) edge[post] (poweron)
        (tsuccess0) edge[post] (xs)
        (tsuccess0) edge[post] (auto)
        (tsuccess0) edge[post] (idle);
        
    \node[transition,minimum size=0.5cm,label=right:\texttt{t_go_to_success_is_arrived_1}] (tsuccess1) at (7.5,1) {};
    \draw[thick] 
        (idle) edge[post] (tsuccess1)
        (is) edge[post] (tsuccess1)
        (auto) edge[post] (tsuccess1)
        (poweron) edge[post] (tsuccess1)
        (tsuccess1) edge[post] (poweron)
        (tsuccess1) edge[post] (xs)
        (tsuccess1) edge[post] (auto)
        (tsuccess1) edge[post] (idle);
    
    \end{tikzpicture} 
    \caption{Start and success transitions of the skill \texttt{go_to} as generated by SkiNet. \texttt{e\_go\_to}, \texttt{i\_go\_to} and \texttt{x\_go\_to\_success\_is\_arrived} are the internal places of the skill, representing the Idle, Running and Ended states respectively. $PowerOn$, $AutoMode$, $Idle$ and $Busy$ are the places representing the states of the resources used with the skill.}
    \label{fig:spot_skill_transits}
\end{figure}

Using the skillset model of Spot\textsuperscript{\tiny\textregistered} shown in Fig \ref{fig:spot_skillset}, the generation process of two transitions, $\tau_{go\_to,start}$ and $\tau_{go\_to,sucess,is\_arrived}$ of the skill \texttt{go_to}, will be presented, and the results shown in Fig. \ref{fig:spot_skill_transits}.

\begin{itemize}
    \item Start transition :
    \begin{equation}
    \begin{aligned}
        \tau_{go\_to,start} = \{
        & \phi = \{( \texttt{lease_status}==AutoMode
        \land \texttt{control_mode}==Idle ) \\
        & \land \texttt{power_status}==PowerOn \}, \\
        & \mathcal{E} = \{\texttt{control_mode}->Busy\}, \\
        & \sigma = (e_{go\_to},i_{go\_to})\}
    \end{aligned}
    \end{equation}
    The guard $\phi$ of this transition is the product of the preconditions \texttt{canmove} and \texttt{ispowered}, while the effects set $\mathcal{E}$ contains the start effect. The set of solutions that satisfies $\phi$ contains only one element:
    \begin{equation}
        X_{\tau} = \{(AutoMode,Idle,PowerOn)\}
    \end{equation}
    The resource states in $x$ will be used to build the input places of the start transition. Because \texttt{power_status} and \texttt{lease_status} are unaffected, the places $PowerOn$ and $AutoMode$ are both inputs and outputs, in order to restitute the tokens.
    The resource \texttt{control_mode} is nominal as it is both guarded and affected.
    The transition also moves the token of the skill state places from the idle place $p^{go\_to}_e$ to the running place $p^{go\_to}_i$.
    In conclusion, only one transition is generated, shown in Fig. \ref{fig:spot_skill_transits}:
    \begin{equation}
        T_{\tau_{go\_to,start}} = \{ \texttt{t_start_go_to} \}
    \end{equation}
        
    \item Success transition :
    \begin{equation}
    \begin{aligned}
        \tau_{go\_to,success\_is\_arrived} = \{
        & \phi = \{ \texttt{lease_status}==AutoMode \land \texttt{power_status}==PowerOn \}, \\
        & \mathcal{E} = \{\texttt{control_mode}->Idle\}, \\ 
        & \sigma = (i_{go\_to},x_{go\_to,success,is\_arrived})\}
    \end{aligned}
    \end{equation}
    The guard $\phi$ is the product of the invariants, here \texttt{is_auto} and \texttt{is_powered}, and the effects set $\mathcal{E}$ contains the success effect. The set of solutions that satisfies $\phi$ again contains only one element:
    \begin{equation}
        X_{\tau} = \{(AutoMode, PowerOn)\}
    \end{equation}
    The resource \texttt{lease_mode} is again unaffected, so the state that solves $\phi$, $AutoMode$, is both input and output of the transitions.
    \texttt{control_mode} is unguarded this time, so two transitions are needed in order to account for all possible states of the resource that can lead to $Idle$, which are $Idle$ and $Busy$. So one transition moves will move the token from $Idle$ to $Idle$, and the other from $Idle$ to $Busy$. Finally, the token of the skill state places is moved from $p^{go\_to}_i$ to the exit place $p^{go\_to}_{x,success,is\_arrived}$. The generated transition are then, as shown in Fig. \ref{fig:spot_skill_transits}:
    \begin{equation}
    \begin{aligned}
        T_{\tau_{go\_to,success\_is\_arrived}} = \{ &\texttt{t_go_to_success_is_arrived_0}, \\ &\texttt{t_go_to_success_is_arrived_1}\}
    \end{aligned}
    \end{equation}

\end{itemize}

\subsection{Initial Marking}

The initial marking of the net follows the initial state of the skillset.
Resources are in their initial state, so only this place has a token, i.e:
\begin{equation}
    r \in \mathcal{R}, p^r_0 = 1, p^r_{i \neq 0} = 0
\end{equation}
Skills are all idle at the initial state, so we have:
\begin{equation}
s \in \mathcal{S}, p^s_e = 1, p^s_i = 0, \forall k, p^s_{x,k} = 0
\end{equation}

Because resources and skills states do not stack, i.e. a resource/skill cannot be twice in the same state at the same time, only one token must be present in a place for any marking. This is called 1-safeness:
\begin{equation}\label{eq:Safeness}
    \forall m \in M, \forall p \in P: m[p] \in \{0,1\}
\end{equation}

\section{Approach Validation}

In this section, the validation of the translation process presented in section 4 will be conducted, using mathematical induction, to prove that the generated Petri net satisfies the state-machine properties of resources and skills. First, the two properties to verify will be presented, and their validity at the initial marking will be proven. Then, the induction for any marking will be developed, before concluding on the validity of the properties.

\subsection{Place invariants of state-machines}

The resources places and skills state places must share only one token at all time. It is critical that the generated transitions will satisfy the two following properties:

\newtheorem{proposition}{Proposition}
\begin{proposition}[Invariants]
For any marking of the net, i.e. $\forall m_n \in M$ the following equations are satisfied:
\begin{equation}\label{eq:ResInv}
    \forall r \in \mathcal{R},\ \sum_{S^r_i \in S^r} m_n[p^r_i] = 1
\end{equation}
\begin{equation}\label{eq:SkInv}
    \forall s \in \mathcal{S},\ m_n[p^s_e] + m_n[p^s_i] + \sum_{k} m_n[p^s_{x,k}] = 1
\end{equation}
\end{proposition}

By extension of this definition, because the set of places $P$ of the generated Petri net is only composed of the resources and skills places, the safeness of the net, Eq. \eqref{eq:Safeness}, is proven if Eq. \eqref{eq:ResInv} and \eqref{eq:SkInv} are satisfied:

\newtheorem{lemma}{Lemma}
\begin{lemma}[Safeness]
    If Eq. \eqref{eq:ResInv} and\eqref{eq:SkInv} are satisfied, then the Petri net is 1-safe, i.e.:
    \begin{equation}
        \forall m \in M, \forall p \in P, m[p] \leq 1
    \end{equation}
\end{lemma}

Let us begin the proof with the initial state.

\selectlanguage{english}
\begin{proof}[Proof by induction - Initial state]

    At $n=0$, $\forall r \in \mathcal{R}$, the state-machine of $r$ is in the initial state $state(r) = S^r_0$, the initial marking of the net is $m_0[p^r_0] = 1$, and $m_0[p^r_{i\neq 0}] = 0$.
    This gives $\sum_{S^r_i \in S^r} m_0[p^r_i] = 1$ and satisfies equation \eqref{eq:ResInv} for any $r \in \mathcal{R}$.

    Additionally, $\forall s \in \mathcal{S}$, all skills are in the idle state $e_s$, therefore $m_0[p^s_e] = 1$, $m_0[p^s_i] = 0$ and $\sum_{k} m_0[p^s_{x,k}] = 0$.
    This gives $m_0[p^s_e] + m_0[p^s_i] + \sum_{k} m_0[p^s_{x,k}] = 1$ and satisfies equation \eqref{eq:SkInv} for any $s \in \mathcal{S}$.
    
    Therefore, the proposition is true for the initial marking $m_0$.
\end{proof}

Now, let us assume that, for a marking $m_n$, properties 1 and 2 are true.
$m_{n+1}$ is the state that follows the firing of any enabled transition $t \in T = ( \bigcup_{\tau} T_\tau )\cup T_{reset}$. First, the reset transitions will be studied, then the event and skill transitions.

\subsection{Reset transitions}
    
Let us prove that for any firing of a reset transition, the properties \eqref{eq:ResInv} and \eqref{eq:SkInv} are still valid in the resulting marking.
    
\begin{proof}[Proof by induction - Reset transitions]
    If $t_{s,reset,k} \in T_{reset}$, a reset transition, is fired, then it will simply transfer the token from one of the termination state place $p^s_{x,k}$ to $p^s_e$, i.e. $^{\bullet}t_{s,reset,k} = \{p^s_{x,k}\}$ and $t_{s,reset,k}^{\bullet} = \{p^s_e\}$. Because equation \eqref{eq:SkInv} is true for $m_n$, we have:
\begin{equation}
    m_n[p^s_{x,k}] = 1,\ m_n[p^s_{x,k' \neq k}] = 0,\ m_n[p^s_e] = 0, m_n[p^s_i] = 0
\end{equation}
    Therefore, firing $t$ will yield:
\begin{equation}
    m_{n+1}[p^s_e] = 1,\ m_{n+1}[p^s_{x,k}] = 0,\  m_{n+1}[p^s_{x,k' \neq k}] = 0,\ m_{n+1}[p^s_i] = 0
\end{equation}
    This holds Eq. \eqref{eq:SkInv}, and because no resource is involved in the firing of $t_{s,reset,k}$, then we are sure that Eq. \eqref{eq:ResInv}, which is true at $m_n$, will be true at $m_{n+1}$.
    We conclude that reset transitions hold both propositions as true when fired.
\end{proof}

\subsection{Event and Skill transitions}

Now, let us prove that for any firing of an event or skill transition, the properties \eqref{eq:ResInv} and \eqref{eq:SkInv} are valid in the resulting marking.

\begin{proof}[Proof by induction - Event and Skill transitions]
    If $t_x \in T_\tau$ is a transition generated from a skillset transition $\tau = (\phi,\mathcal{E},\sigma)$, then we have:
\begin{itemize}
    \item $^{\bullet}t_x = \{p^s_1,p^r_n,...\}$ the input places composed of the state place $p^s_1$ and a unique resource state place $p^r_n$, corresponding to $state(r)$ at the marking $m_n$, for each resource $r \in \mathcal{R}(\tau)$.
    \item $t_x^{\bullet} = \{p^s_2, p^r_{n+1}, ...\}$ the output places composed of the state place $p^s_2$ and a unique resource state place $p^r_{n+1}$ at the marking $m_{n+1}$ for each $r \in \mathcal{R}(\tau)$.
\end{itemize}
    Because Eq. \eqref{eq:ResInv} is true, for any $r \in \mathcal{R}(\tau)$, a single token is present in $r$ at the state place $p^r_n$ at the marking $m_n$, and the firing of $t_x$ transfers the token to the state place $p^r_{n+1}$ at the following marking $m_{n+1}$, i.e.:
    
\begin{equation}
    \forall r \in \mathcal{R}(\tau),\ m_n[p^r_n] = 1,\ m_n[p^r_{i \neq n}] = 0
\end{equation}
    And firing $t$ leads to:
\begin{equation}
    \forall r \in \mathcal{R}(\tau),\ m_{n+1}[p^r_{n+1}] = 1,\ m_{n+1}[p^r_n] = 0,\ m_{n+1}[p^r_j] = 0,\ j \not \in \{n,n+1\}
\end{equation}
    Giving: $\forall r \in \mathcal{R},\ \sum_{S^r_{i} \in S^r} m_{n+1}[p^r_i] = 1$, so equation \eqref{eq:ResInv} is satisfied.
\end{proof}

    N.B.: it is possible that the input state and destination places $p^r_n$ and $p^r_{n+1}$ are the same, since $S^r_i \in {S^r_i}^\bullet, \forall S^r_i \in S^r$, as defined in Def. \ref{def:DefResources}. In that case, the firing of $t$ will take the token from $p^r_n$ and restitute it, so if Eq. \eqref{eq:ResInv} was true at $m_n$, then it will be true at $m_{n+1}$.

    For Eq. \eqref{eq:SkInv} , because $t$ transfers the token from $p^s_1$ to $p^s_2$, for any skill $s \in \mathcal{S}$ , and because Eq. \eqref{eq:SkInv}  is true at $m_n$, then we can directly conclude that Eq. \eqref{eq:SkInv} will hold upon the firing of $t$ as both $x_s$ and $y_s$ only contain one place among the internal places of $s$. Additionally, since $\sigma = \varnothing$ for events, then no skill state place is involved in the firing of $t_x$, therefore Eq. \eqref{eq:SkInv} is automatically verified at $m_{n+1}$.
    
    Therefore, both properties \eqref{eq:ResInv} and \eqref{eq:SkInv} hold upon the firing of any $t \in T_\tau$. Finally, because the set of all transitions of the generated Petri net $T$ is the union of the set of reset transitions $T_{reset}$ and the events and skills transitions $\bigcup_{\tau} T_\tau$, then all transitions of $T$ will hold the proposition.
    
\subsection{Conclusion}

    Having proven that properties 1 and 2 are true at the initial marking $m_0$ of the generated Petri net, and that for any $m_n \in M$, the firing of any transition $t \in T$ holds the properties from $m_n$ to $m_{n+1}$ the marking reached by firing $t$, we can conclude that, for any marking $m$, the proposition is true.
    
    This means that our generated Petri net respects the state-machine properties of resources and skills internal places, and that its behavior reflects correctly the behavior of the skillset as it would run on the autonomous system it models. We can now use the generated Petri net to do $a\ priori$ verification by using Petri net analysis tools available in the litterature. 
    
\section{Model Verification}

Now that the net generation aspect of SkiNet has been presented and validated, the possible model-checking thanks to the net will be presented in this section. SkiNet offers a simple interface to execute all or part of the following verification steps, without the need for the users to use the model-checking tools directly. However, it will still display the formulas used, for transparency, but also to guide the users if they wish to check their own properties with the same model-checking tools. The corresponding SkiNet verification name for each step is indicated in the subsection name, between quotation marks, and is tested on the \texttt{spot} skillset in Fig. \ref{fig:spot_skillset}. For reference, the computation time of the Petri net generation and its verification take less than a minute. Adding more resources and skills increases the calculation time, but so far no tested skillset was found to exceed a few minutes for the whole process. For flexibility purpose, options were added to remove events or exit places (tokens are directly moved to the entry place of skills upon firing an exit transition, and reset transitions are removed), to increase model-checking capabilities, computational time and overall scalability.

\subsection{Kripke Structure of the skillset net}

In order to perform model-checking using the tools Selt (LTL) and Muse (mu-calculus, CTL) of the Tina toolbox, we first need to convert the generated Petri net into a Kripke structure, on which temporal logic operations can be performed. Liu \cite{Liu2012} summarizes the Kripke structure definition, generated, and later used as input, by the Tina toolbox. More informations and instructions can be found on the Tina toolbox website \cite{TinaWebsite}.

\subsection{Petri net Deadlocks - "dead"}

Deadlocks are markings where no transition is fireable. In this context, it means that no actions from skills or events can be done, and no reset transition is available. This can happen purposefully, for instance if a resource represents the battery state of the robot, and upon entering a critical state, all skills and events are blocked and nothing can happen anymore. However, it can also be the result of an error in the specifications of the skillset, so it's important to know what sequence of transitions led to the dead marking. 
In the tool Selt, the predicate \emph{dead} has already been implemented to check if a state is a deadlock, therefore the LTL formula is to check for deadlocks in the net is "we never have a deadlock". Using the temporal operator "A" (Always):
\begin{equation}
    A \lnot dead
\end{equation}\label{eq:LTLdead}
The program will either return $True$ if no path leads to a deadlock, $False$ if there is one, with a counter-example showing a sequence of transitions leading to the dead state.
In the example skillset \texttt{spot} from Fig. \ref{fig:spot_skillset}, no deadlock was found, as the events can be fired infinitely, so there is always an event transition enabled.

\subsection{Petri net Liveness - "live"}

The non-dead transitions are searched. If a transition is never fireable, it is considered dead. Dead transitions appear almost in every case of net generation, and are due to the transition generation algorithm presented in section 4.2, which can find solutions that will never happen.

For the generated Petri net, a transition is not dead if it can be fired at least once, which can be checked with the formula:
\begin{equation}
    \forall t \in T : A \lnot t
\end{equation}
If the formula is true, then the transition is dead. SkiNet will return the list of all dead transitions when running the "live" option.
Looking at the \texttt{spot} skillset and the example success transition in Fig. \ref{fig:spot_skill_transits}, \texttt{t_go_to_success_is_arrived_1} was found to be a dead transition. This is because the resource \texttt{control_mode} is never in the state $Idle$ during the execution of the skill \texttt{go_to_body}.
This verification step can become quite time consuming for Petri nets with a large amount of markings, but the \texttt{spot} example is ran in only a few seconds as it is very simple.

\subsection{Skillset Invariants verification - "safe"}

In section 5, it was verified that the transitions generated respect the resources and skills places invariants, i.e. only one token exist at any marking in the same resource/skill places.
In order to verify Eq. \eqref{eq:ResInv} and \eqref{eq:SkInv} using LTL, the following formula can be used for resources, which verifies that the sum of tokens among the places of the resource is always one:
\begin{equation}
    \forall r \in \mathcal{R} : A ( \sum_{S^r_i \in S^r} p^r_i = 1)
\end{equation}
And for skills:
\begin{equation}
    \forall s \in \mathcal{S} : A ( p^s_e + p^s_i + \sum_k p^s_{x,k}) = 1)
\end{equation}

Finally, the safeness of the net presented in Eq. \eqref{eq:Safeness} can also be checked using LTL, i.e. no place has more than one token for any marking:
\begin{equation}
    \forall p \in P : A \lnot( p >= 2)
\end{equation}

These three LTL formulas were mostly used during the development of SkiNet to verify that the code would generate a proper net as defined in section 4 and 5.

\subsection{Skills and Skillset deadlocks - "deadskill" and "deadset"}

A property specific to the skillset architecture that can be verified is whether a skill, two concurring skills, or all the skills, are always alive or not. In other terms: is there a configuration in the skillset that can render a skill forever unusable. Here, LTL cannot be used, as these properties need to be checked for all paths, from any state. Therefore, mu-calculus and CTL are used with the Kripke structure, through the Muse tool of Tina.

The CTL property to check whether a skill is always eventually activatable is "For all paths globally, there is some path where finally $p^s_i$".
\begin{equation}
    AG EF p^s_i, s \in S
\end{equation}
This property either returns all the states if it is true, and none if there exist a single state where it is false. But because it returns no state, it does not provide sufficient feedback for the user.
Therefore, in order to know the specific transition sequences leading to a state that would make the skill forever unactivatable, the property "For all paths finally, there are no paths where finally $p^s_i$" is used, so that the Muse tool returns exploitable counter-examples:
\begin{equation}
    \lnot AF EF p^s_i, s \in S
\end{equation}
The states returned are then stored and the transition paths leading to these states can be obtained using the Pathto tool of Tina and the Kripke structure. For now, only the first of these states is returned by SkiNet. If the list of states is empty, then the skill can always be activated.
It is possible to check for all skills at once with the property:
\begin{equation}
    \lnot AF EF (\sum_{s\in S} p^s_i)
\end{equation}

In our example \texttt{spot}, an error was found in the skillset using these properties. SkiNet pointed out that, when executing the \texttt{go_to} skill, the failure of invariants \texttt{is_auto} and \texttt{is_powered} would block all the other skills from executing as the resource \texttt{control_mode} would not go back to the $Idle$ state upon exiting \texttt{go_to}, thus blocking the precondition \texttt{canmove} of the skills. This was fixed by adding a failure effect \texttt{control_mode -> Idle} to the invariants (not shown in the Fig. \ref{fig:spot_skillset}).

\section{Conclusion}

The work presented in this paper aim to propose a method to verify the behavior of a software control architecture for critical autonomous systems. A tool was developed, called SkiNet, which generates a Petri net that models the behavior of the system as specified in the skillset model. Using the generated Petri net, SkiNet allows the users to check their skillset model by using model-checking. Basic properties such as Petri net deadlocks or transition liveness can be checked, as well as more skillset-specific properties such as skill liveness/deadlocks. Because the skillset-architecture upon which this work is based on is also a work-in-progress, future works will focus on proving the equivalence between the skillset semantic during runtime and the generated Petri net semantic, once the former is well established. Additionally, with always the same focus of making formal verification accessible to all actors of a robotic project, regardless of their background, two extensions of SkiNet are under progres. SkiNet Mission will allow users to create and verify the feasability of their missions, using Petri nets and model-checking, and SkiNet Live will offer users a view of the transition firing in the Petri net, synchronized with the skillset as it runs on the robotic system, for runtime analysis. The checking of properties during runtime is also being considered for the latter.
Among the difficulties encountered, the complexity of the state space analysis of the generated Petri net is the biggest issue, as the addition of resources and skills to a skillset make the Kripke structure grow exponentially, thus increasing model-checking complexity. This is especially difficult to handle when in the iterative phase of the model definition. Petri net reduction algorithms may be considered as future work if this issue remains or makes the tool inoperable in a realistic development process of an autonomous system.


\bibliographystyle{eptcs}
\bibliography{generic}
\end{document}